%% file: ms.tex
\documentclass{iau}
\usepackage{graphicx}
\usepackage{epsfig}
\usepackage{multirow}
\usepackage{natbib}
\usepackage{deluxetable}

\title[IAUS 314.~~The Dustiest Main Sequence Stars] 
{Can the dustiest main sequence stars tell us about the rocky planet formation process?}

\author[C.\ Melis]   
{Carl Melis$^1$}

\affiliation{$^1$Center for Astrophysics and Space Sciences, University of California, San Diego, California 92093-0424, USA\\email: {\tt cmelis@ucsd.edu}}

\pubyear{2015}
\volume{314}  
\pagerange{119--126}
\jname{Young Stars \& Planets Near the Sun}
\editors{J. H. Kastner, B. Stelzer, \& S. A. Metchev, eds.}

\input{astmac}

\begin{document}

\maketitle

\begin{abstract}
Main sequence stars hosting extreme quantities of inner planetary system debris 
are likely experiencing transient dust production events. The nature of these events,
if they can be unambiguously attributed to a single process, can potentially inform us
on the formation and/or early evolution of rocky Earth-like planets. In this contribution I 
examine some of the dustiest main sequence stars known and
three processes that may be capable of reproducing their observed properties.
Through this activity I also make an estimate for the likelihood of an A-type star
to have an asteroid belt-like planetesimal population.
\keywords{planets and satellites: formation, circumstellar matter, planetary systems}
\end{abstract}

\firstsection 
\section{Introduction}

Understanding how rocky terrestrial planets form and evolve is paramount to the 
question of how life forms on planets and whether or not other sentient beings can 
exist in the universe. Currently the best means of identifying stars undergoing terrestrial 
planet formation or collisional evolution events is discovery and characterization of dusty debris 
disks that orbit in the inner planetary system of their host star 
(e.g., \citealt{kenyon06}; \citealt{melis10}; \citealt{jackson12}; \citealt{meng14,meng15}; \citealt{genda15}).
Dust at a distance of $\sim$1\,AU from a Sun-like star will be heated
to Earth-like temperatures, $\sim$300 K, and will emit in the mid-infrared,
thus most surveys for inner planetary system dust look for excess emission
at wavelengths of 10-20\,$\mu$m. 

Investigations into how mid-infrared excess stars provide insight into the
rocky planet formation process are being actively pursued by several groups.
Statistical works conducted to date
typically do not discriminate on how dusty a given debris disk system 
is when including it in their analysis (e.g., \citealt{meyer08}; \citealt{jackson12};
\citealt{genda15}).
It is worth noting that it need not be the case that all mid-infrared excesses are 
linked to rocky planet formation events. Indeed, another hypothesis
is that weaker excesses can be attributed to dust produced through collisional evolution
of asteroid belt-analogs. Such
a suggestion seems quite appropriate for the \citet{meyer08} sample which covers an
age range of 0.3-3.0\,Gyr where one would not expect rocky planet formation collisional events to 
be occurring. As such, one might interpret the \citet{meyer08} statistics as telling us
that 19-32\% or up to 62\% of Sun-like stars host
asteroid belt analogs. Under this banner, I
estimate the occurrence rate for asteroid belts around A-type
stars.

Examination of the incidence rate of such systems at young ages (10-30\,Myr old), 
when collisional cascade dust production is at its height \citep[e.g.,][]{wyatt08}, 
should enable a reasonable estimate of the fraction of stars that
host asteroid-belt analogs.
To ensure a robust estimate, it is important to use a statistical sample of stars observed
in a blind survey where it is clear that the mid-infrared excess emission originates from inner 
planetary system material. 
\citet{morales09} provide just such a data set 
by combining {\it Spitzer} MIPS and IRS measurements for 
relatively luminous stars. Their data set, which probes stars of spectral type A and late-B that were
previously determined to have excess emission at 24\,$\mu$m, shows that
10 out of 14 (or 71\%) of 24 and 70\,$\mu$m detected stars have blackbody-fit dust 
temperatures $\gtrsim$200\,K (only those stars with single-temperature
blackbody fits that are detected at 24 and 70\,$\mu$m are selected
to ensure that the dust temperature $-$ and hence physical location $-$ is well
constrained). All of these 10 sources have a fractional infrared 
luminosity that is $<$2$\times$10$^{-4}$
and thus can be interpreted as hosting
active asteroid belt analogs (similar to the conclusion reached by \citealt{morales09}). 
\citet{rieke05} and \citet{su06} present unbiased 24\,$\mu$m excess statistics, and each find that
roughly 40-50\% of A-type stars with ages between 10-30\,Myr host 24\,$\mu$m
excess emission. When combined with the finding from \citet{morales09} that
$\approx$71\% of A-type stars with 24\,$\mu$m excess emission have dust
in their inner planetary system, it is found that at least 33\% of A-type stars
with age in the range of 10-30\,Myr should host an active planetesimal belt
in their inner planetary system. This occurrence rate is in good agreement
with the statistics of polluted white dwarf stars discussed
by \citet{zuckerman03} and \citet{zuckerman10}, indicating that such asteroid 
belt-analogs go on to eventually pollute their dead stars' atmospheres.

Rather than working with all mid-infrared excess main sequence stars, one might instead opt
to examine only those hosting extreme quantities of inner planetary system material as
they are likely experiencing transient dust production events.
To date, only a handful of main sequence stars are known to have sufficiently high levels of 
mid-infrared excess emission (and hence inner planetary system dust) that points to an
unambiguous origin in transient processes that may be relevant to 
terrestrial planet formation or evolution
(e.g., \citealt{song05}, \citealt{rhee07b}, \citealt{rhee08}, 
\citealt{zuckerman12}, \citealt{melis12}, \citealt{olofsson12},
\citealt{schneider13}, \citealt{melis13}, \citealt{kennedy14}).
\citet{melis14} describe how to sort inner planetary system debris
disks into two general categories: those that result from
collisional grinding down of a population of rocky planetesimals
(an active asteroid belt-analog) and those that require
a transient dust production event (stochastic or transient collisions involving
planetary-scale objects). They begin with the assumption that 
all dust disks are the product of collisions of numerous
small rocky bodies in an active planetesimal belt. It is then estimated
what total mass of parent bodies $-$ $M_{PB}$ $-$ is necessary to reproduce
the observed parameters for the small dust grains and compare $M_{PB}$ to
the model simulations of \citet{kenyon06} for the formation of terrestrial
planets to see if this amount of mass should have coalesced to form 
planetary embryos or planets. Should the indicated mass of
parent bodies be sufficient to expect the formation of planetary-scale
bodies ($M_{PB}$$\gtrsim$1\,M$_{Earth}$), then it is concluded
that the active planetesimal belt hypothesis is false 
(as any planetesimal population should have developed into planetary embryos
or planets) and that the observed dust results from transient dust-production
involving large rocky objects.

\begin{figure}
 \centering
 \begin{minipage}[h!]{120mm}
 \includegraphics[width=120mm]{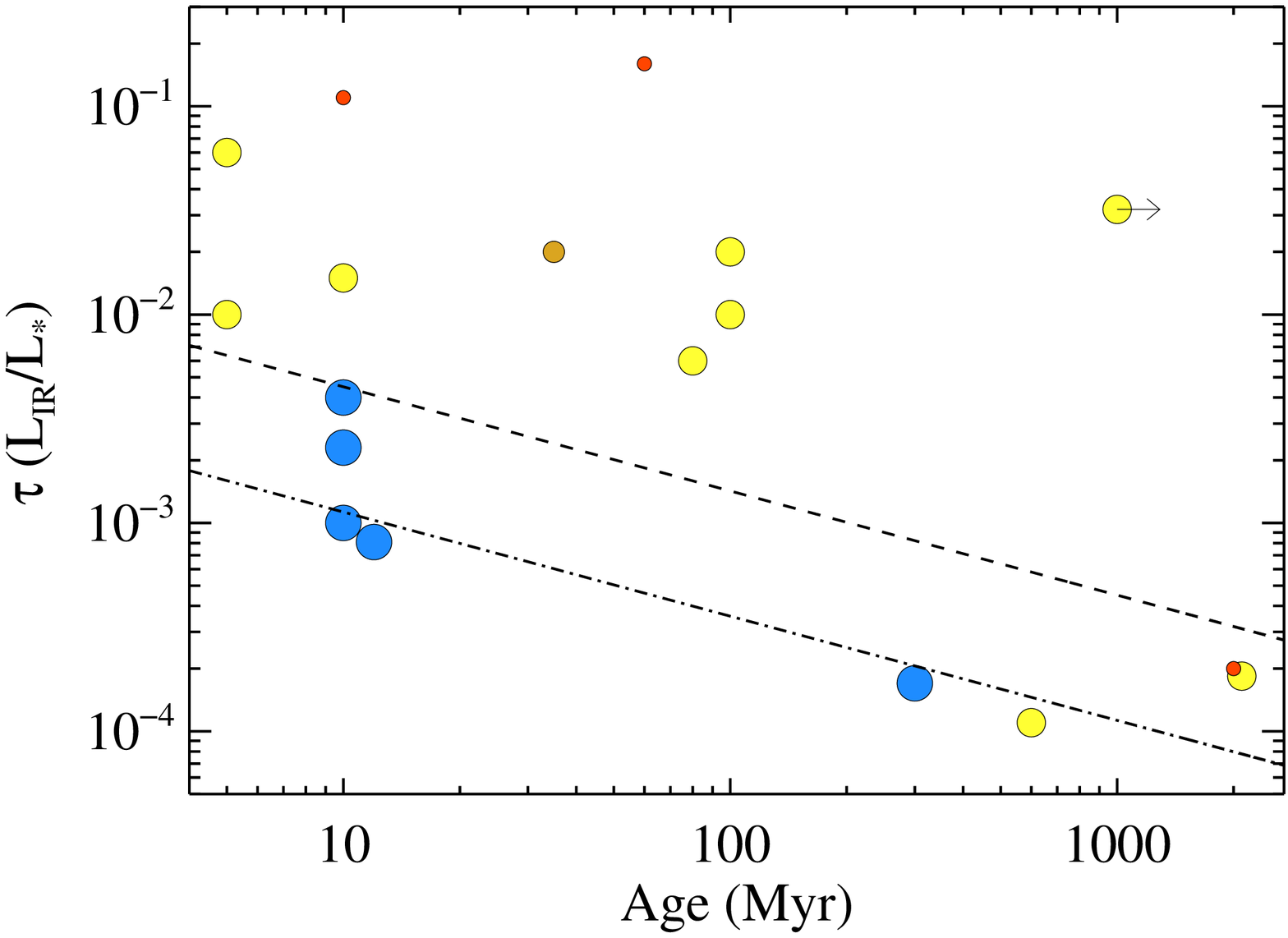}
\end{minipage} 
\\*
\begin{minipage}[h!]{120mm}
 \includegraphics[width=120mm]{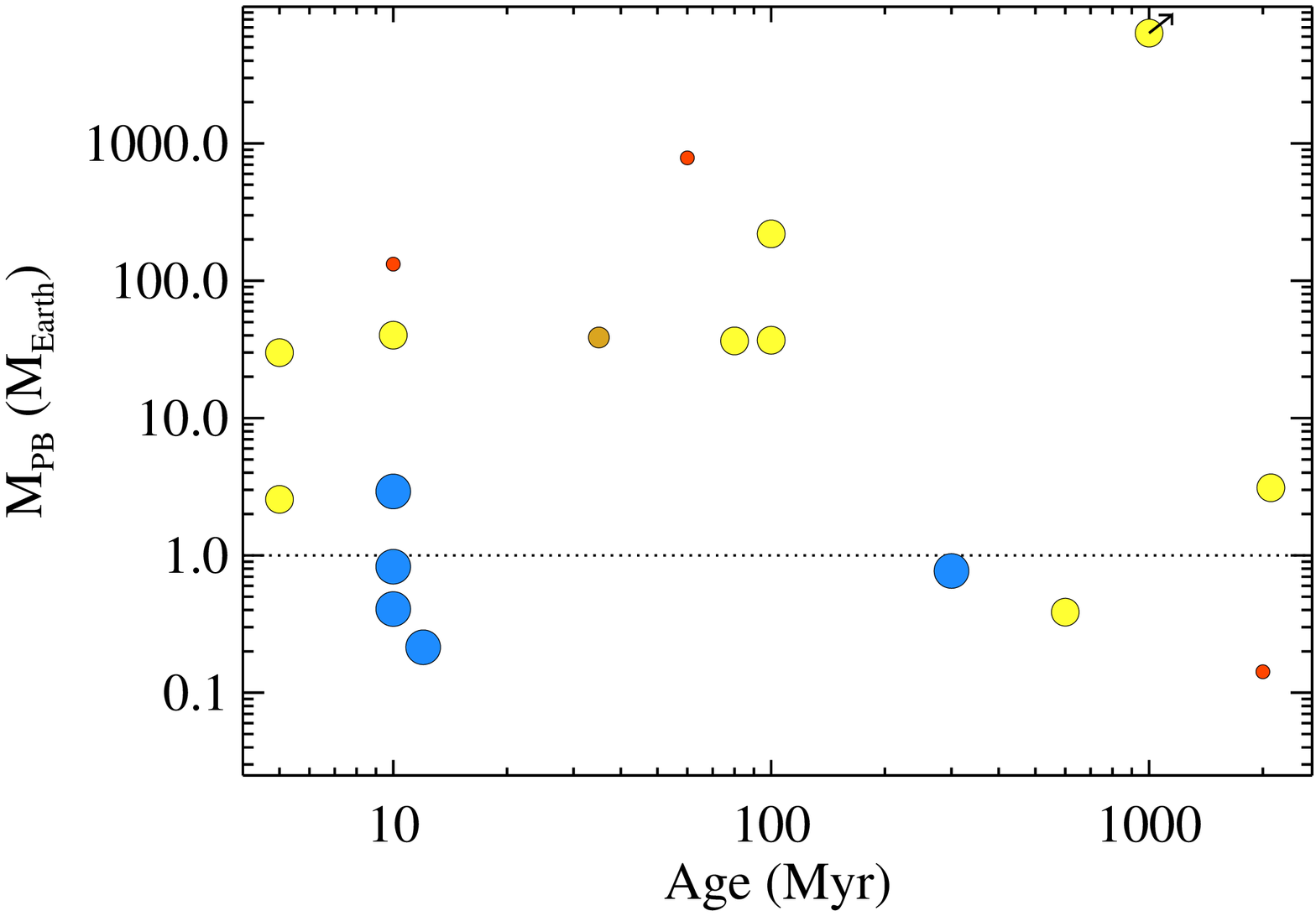}
\end{minipage} 
\caption{\label{figwarmdust} {\it Both Panels:} 
               Stars hosting large quantities of inner planetary system material
               likely generated in transient dust production events
               (M$_{PB}$$\gtrsim$1\,M$_{Earth}$; see \citealt{melis14}).
               Objects down to 0.1\,M$_{Earth}$ are still considered
               as candidates but could be explained by collisional grinding down of an active
               planetesimal belt; some representative systems are shown. 
               Sources are drawn from \citet{melis10}, \citet{melis13},
               and references therein.
               Small, red circles are K-type stars. The medium-small, gold circle is
               a G-type star. Medium-large, yellow circles are F-type stars. Large,
               blue circles are A-type stars. The object with a limit arrow and age
               of 1000\,Myr is BD+20~307 as its age is a lower limit.
               {\it Upper Panel:} The dashed line represents $\tau$ necessary to realize
               a parent body mass of 1\,M$_{Earth}$ for a star of solar mass, radius, 
               and temperature that is orbited by 300\,K dust. The dash-dotted line represents
               the $\tau$ necessary to realize a parent body mass of 1\,M$_{Earth}$ for
               a star of 1.7\,R$_{\odot}$, 2.0\,M$_{\odot}$, and 8180\,K effective temperature
               that is orbited by 300\,K dust.
               {\it Lower Panel:} The
               dotted line represents the cutoff above which the
               observed dust is expected to be the result of transient processes
               (see \citealt{melis14}).}
\end{figure}

\section{Extreme debris disk systems}
\label{sectpfs}

All main sequence stars currently known and confirmed
to be hosting substantial quantities of inner planetary system dust $-$
and thus likely to be undergoing transient dust production events
(M$_{PB}$$\gtrsim$1\,M$_{Earth}$) $-$ are presented in 
Figure \ref{figwarmdust}.
Examination of the properties of these systems reveals that
those with early-type host stars are mostly observed 
when the star is $\sim$10\,Myr old.
By contrast, this epoch extends later for Sun-like stars with observed specimens 
typically having ages $\leq$100\,Myr. The incidence rate of the dustiest Sun-like stars,
determined by \citet{melis10} to be roughly 1 in 300 for the age range of 30 to 100\,Myr,
is comparable to that for early-type stars of approximately 1 in 200 for the age
range of 10 to 20\,Myr (Melis et al.\ in preparation). If all stars undergo
an exceptionally dusty phase of inner planetary system evolution, then the observable
signature of it should persist for about 1.5$\times$10$^5$~years for Sun-like stars 
\citep{melis10} and $\sim$7$\times$10$^4$~years for early-type stars
(Melis et al.\ in preparation).

To be consistent with the data presented in
Figure \ref{figwarmdust}, any model should reproduce the high level of dust
observed around each star, the dominant timescale in which this dust
is observed, and the observed incidence rate.
Three possibilities are presented below.

\subsection{Giant Impacts}
\label{secgimp}

\citet{rhee08} and \citet{melis10} interpret the incidence rate
of the dustiest main sequence stars with the aid of
colliding planetary embryo models
developed in \citet{agnor99}, \citet{agnor04}, and \citet{asphaug06}. 
In brief, following a giant impact
of two rocky planetary embryos or planets, fragmented debris covering a range
of sizes is released. 
The collision time for the smallest dust particles produced in the ensuing collisional 
cascade will be about 1 year
divided by 80$\times$L$_{\rm IR}$/L$_{\rm bol}$ (\citealt{melis14} and references therein).
If the largest initial fragments
ejected during a giant impact have radii $\sim$100\,m (see work by 
Agnor, Asphaug, and colleagues $-$ note that this is in contrast to the suggestion by
\citealt{jackson12} and \citealt{genda15} who suggest ejecta up to sizes
of 500\,km in diameter), then their collisional
lifetime is $\sim$5$\times$10$^4$~years for Sun-like stars and
$\sim$2$\times$10$^4$~years for early-type stars (see \citealt{melis10}). 
However, the formation of rocky terrestrial planets involves more
than a single giant impact event (e.g., \citealt{stewart12}).
Through the analysis of \citet{stewart12} and the simulations referenced therein
it is found that a typical terrestrial planet will experience $\sim$6
giant impact-type events before it is fully formed. If half of these result in
the production of large monolith fragments (as suggested
by \citealt{stewart12}), then the model-estimated observable transient dust 
lifetime could be a few times more than the above estimates
and hence in reasonably good agreement with the values given at the beginning
of this section. If accurate, this suggests that on average
there is one terrestrial planet for every star of 
Solar-mass and above (this does not necessarily mean all stars have terrestrial
planets, as some may have multiple and others none). This is in good agreement
with the recent determination of the prevalence of rocky bodies orbiting other stars
from the {\it Kepler} data set \citep[e.g.,][]{burke15}.

\subsection{Leftover Planetesimals}

Radiometric age-dating of Solar system rocky bodies provides a detailed account
of the impact history in the inner parts of our planetary system. Study of impact-reset
$^{40}$Ar$-$$^{39}$Ar ages in lunar samples and meteorites shows the well-known
increase in impactor flux between 3.4 and 4.1\,Gyr ago (usually attributed to a
Late-Heavy Bombardment phase), but meteorite samples alone also show a
spike of reset activity around $\approx$4.5\,Gyr ago (e.g., see Figure 1 of \citealt{marchi13}
and references therein). As discussed in \citet{marchi13},
these radiometric reset events are often attributed to cooling of the parent body, but
substantial evidence suggests that impacts are also important in generating
this spike. Impacts in this time frame (ages of 10-100\,Myr for the Sun), 
if indeed sufficiently energetic to reset radiometric
geochronometers, likely would have produced significant quantities of dust. 
Such impacts could in theory be capable of
explaining the data presented in 
Figure \ref{figwarmdust}.

\citet{marchi13} postulate that impactors capable of producing the 
$\approx$4.5\,Gyr old $^{40}$Ar$-$$^{39}$Ar spike could come from a 
class of ``leftover planetesimals'' excited to
high orbital eccentricity and inclinations by forming rocky protoplanets (see also \citealt{bottke07}).
\citet{bottke07} show that a scattered planetesimal population
would rapidly deplete, with $\approx$90\% of the sample being
removed from the planetary system by the time its host star is $\sim$100\,Myr old. The
implication is that dust production from this depletion would be dominated by 
planetesimal-asteroid collisions, although some material should impact the rocky planetary
bodies that expelled it. 
Large rocky planetary bodies are necessary in generating the
dust, although they will typically lack an immediate connection to it. As an interesting
aside, it is prudent to wonder if such objects could also be responsible for late
accretion or veneer \citep[e.g.,][]{day12,sch12}.

\subsection{Dynamical Instability}

Solar system formation models
\citep[e.g., the Nice model;][]{gomes05,morbidelli05,tsiganis05}
predict a phase of outer planetary system reconfiguration wherein the inner planetary 
system experienced an enhanced influx of planetesimals. This ``Late Heavy Bombardment'' $-$
which occurred $\approx$600\,Myr after the Sun was born $-$
has observational evidence in the cratering record of various inner Solar system rocky objects. It is
enticing to consider a similar phase of dynamical instability as capable of driving the exceptionally
dusty states of the stars considered in Figure \ref{figwarmdust}
(indeed, \citealt{fujiwara09,fujiwara12} advance dynamical instabilities to explain the
dusty stars they study).

Some studies have shown that it is possible for dynamical instabilities to produce
the observed incidence rate of the dustiest main sequence stars (e.g., \citealt{bonsor13,bonsor14} 
and references therein). These simulations also suggest the incidence rate of dusty systems resulting
from instabilities and the level of dust being produced by them tend to decrease
with increasing stellar age, again in reasonable agreement with the data in Figure \ref{figwarmdust}.
But, it is not clear if the level of dust observed can be reproduced as instability models have not yet
tracked collisional and dynamical evolution in tandem. 
New codes should allow coupled simulations
and direct tests of these models through predictions for direct observables
(e.g., \citealt{stark09}; \citealt{kral13}).

The implication of this model should it be appropriate 
is that these systems would host giant planets and planetesimal belts (typically in the
outer planetary system) that
are experiencing dramatic reconfigurations; rocky Earth-like planets are not immediately necessary.

\section{Conclusion}

There is strong evidence that transient dust production events
occur in the inner planetary systems of other stars. Examination of the dustiest main
sequence stars as a subset of the debris disk population provides a safe 
route to selecting those systems where it is clear transient events are occurring
and thus to indirectly probing terrestrial planet formation and evolution through 
the statistics of infrared excess stars. 
However, more observations, theory development, and simulations are necessary
before we can robustly attach these systems to specific phases of planetary system formation
or evolution.

\begin{discussion}

\end{discussion}

\end{document}

%% file: astmac.tex
\usepackage{color}
\usepackage{graphicx}

\newcommand{\mnras}{MNRAS}

\newcommand{\apj}{ApJ}

\newcommand{\apjl}{ApJL}
\newcommand{\aap}{A\&A}
\newcommand{\araa}{ARAA}
\newcommand{\aj}{AJ}
\newcommand{\nat}{Nature}

\newcommand{\icarus}{Icarus}

%% file: ms.bbl
\begin{thebibliography}{41}
\expandafter\ifx\csname natexlab\endcsname\relax\def\natexlab#1{#1}\fi

\bibitem[\protect\astroncite{{Agnor} \& {Asphaug}}{2004}]{agnor04}
{Agnor}, C. \& {Asphaug}, E. 2004, {\em \apjl\/}, {\bf 613}, L157

\bibitem[\protect\astroncite{{Agnor} {\em et~al.\/}}{1999}]{agnor99}
{Agnor}, C.~B., {Canup}, R.~M., \& {Levison}, H.~F. 1999, {\em Icarus\/}, {\bf
  142}, 219

\bibitem[\protect\astroncite{{Asphaug} {\em et~al.\/}}{2006}]{asphaug06}
{Asphaug}, E., {Agnor}, C.~B., \& {Williams}, Q. 2006, {\em \nat\/}, {\bf 439},
  155

\bibitem[\protect\astroncite{{Bonsor} {\em et~al.\/}}{2013}]{bonsor13}
{Bonsor}, A., {Raymond}, S.~N., \& {Augereau}, J.-C. 2013, {\em \mnras\/}, {\bf
  433}, 2938

\bibitem[\protect\astroncite{{Bonsor} {\em et~al.\/}}{2014}]{bonsor14}
{Bonsor}, A., {Raymond}, S.~N., {Augereau}, J.-C., \& {Ormel}, C.~W. 2014, {\em
  \mnras\/}, {\bf 441}, 2380

\bibitem[\protect\astroncite{{Bottke} {\em et~al.\/}}{2007}]{bottke07}
{Bottke}, W.~F., {Levison}, H.~F., {Nesvorn{\'y}}, D., \& {Dones}, L. 2007,
  {\em \icarus\/}, {\bf 190}, 203

\bibitem[\protect\astroncite{{Burke} {\em et~al.\/}}{2015}]{burke15}
{Burke}, C.~J., {\em et~al.\/} 2015, {\em \apj\/}, {\bf 809}, 8

\bibitem[\protect\astroncite{{Day} {\em et~al.\/}}{2012}]{day12}
{Day}, J.~M.~D., {Walker}, R.~J., {Qin}, L., \& {Rumble}, III, D. 2012, {\em
  Nature Geoscience\/}, {\bf 5}, 614

\bibitem[\protect\astroncite{{Fujiwara} {\em et~al.\/}}{2009}]{fujiwara09}
{Fujiwara}, H., {\em et~al.\/} 2009, {\em \apjl\/}, {\bf 695}, L88

\bibitem[\protect\astroncite{{Fujiwara} {\em et~al.\/}}{2012}]{fujiwara12}
--- 2012, {\em \apjl\/}, {\bf 749}, L29

\bibitem[\protect\astroncite{{Genda} {\em et~al.\/}}{2015}]{genda15}
{Genda}, H., {Kobayashi}, H., \& {Kokubo}, E. 2015, {\em ArXiv e-prints\/}

\bibitem[\protect\astroncite{{Gomes} {\em et~al.\/}}{2005}]{gomes05}
{Gomes}, R., {Levison}, H.~F., {Tsiganis}, K., \& {Morbidelli}, A. 2005, {\em
  \nat\/}, {\bf 435}, 466

\bibitem[\protect\astroncite{{Jackson} \& {Wyatt}}{2012}]{jackson12}
{Jackson}, A.~P. \& {Wyatt}, M.~C. 2012, {\em \mnras\/}, {\bf 425}, 657

\bibitem[\protect\astroncite{{Kennedy} \& {Wyatt}}{2014}]{kennedy14}
{Kennedy}, G.~M. \& {Wyatt}, M.~C. 2014, {\em \mnras\/}, {\bf 444}, 3164

\bibitem[\protect\astroncite{{Kenyon} \& {Bromley}}{2006}]{kenyon06}
{Kenyon}, S.~J. \& {Bromley}, B.~C. 2006, {\em \aj\/}, {\bf 131}, 1837

\bibitem[\protect\astroncite{{Kral} {\em et~al.\/}}{2013}]{kral13}
{Kral}, Q., {Th{\'e}bault}, P., \& {Charnoz}, S. 2013, {\em \aap\/}, {\bf 558},
  A121

\bibitem[\protect\astroncite{{Marchi} {\em et~al.\/}}{2013}]{marchi13}
{Marchi}, S., {\em et~al.\/} 2013, {\em Nature Geoscience\/}, {\bf 6}, 303

\bibitem[\protect\astroncite{{Melis} {\em et~al.\/}}{2010}]{melis10}
{Melis}, C., {Zuckerman}, B., {Rhee}, J.~H., \& {Song}, I. 2010, {\em \apjl\/},
  {\bf 717}, L57

\bibitem[\protect\astroncite{{Melis} {\em et~al.\/}}{2012}]{melis12}
{Melis}, C., {Zuckerman}, B., {Rhee}, J.~H., {Song}, I., {Murphy}, S.~J., \&
  {Bessell}, M.~S. 2012, {\em \nat\/}, {\bf 487}, 74

\bibitem[\protect\astroncite{{Melis} {\em et~al.\/}}{2013}]{melis13}
--- 2013, {\em \apj\/}, {\bf 778}, 12

\bibitem[\protect\astroncite{{Melis} {\em et~al.\/}}{2014}]{melis14}
--- 2014, in {\em IAU Symposium\/}, edited by N.~{Haghighipour}, vol. 293 of
  {\em IAU Symposium\/},  273--277

\bibitem[\protect\astroncite{{Meng} {\em et~al.\/}}{2014}]{meng14}
{Meng}, H.~Y.~A., {\em et~al.\/} 2014, {\em Science\/}, {\bf 345}, 1032

\bibitem[\protect\astroncite{{Meng} {\em et~al.\/}}{2015}]{meng15}
--- 2015, {\em \apj\/}, {\bf 805}, 77

\bibitem[\protect\astroncite{{Meyer} {\em et~al.\/}}{2008}]{meyer08}
{Meyer}, M.~R., {\em et~al.\/} 2008, {\em \apjl\/}, {\bf 673}, L181

\bibitem[\protect\astroncite{{Morales} {\em et~al.\/}}{2009}]{morales09}
{Morales}, F.~Y., {\em et~al.\/} 2009, {\em \apj\/}, {\bf 699}, 1067

\bibitem[\protect\astroncite{{Morbidelli} {\em et~al.\/}}{2005}]{morbidelli05}
{Morbidelli}, A., {Levison}, H.~F., {Tsiganis}, K., \& {Gomes}, R. 2005, {\em
  \nat\/}, {\bf 435}, 462

\bibitem[\protect\astroncite{{Olofsson} {\em et~al.\/}}{2012}]{olofsson12}
{Olofsson}, J., {Juh{\'a}sz}, A., {Henning}, T., {Mutschke}, H., {Tamanai}, A.,
  {Mo{\'o}r}, A., \& {{\'A}brah{\'a}m}, P. 2012, {\em \aap\/}, {\bf 542}, A90

\bibitem[\protect\astroncite{{Rhee} {\em et~al.\/}}{2007}]{rhee07b}
{Rhee}, J.~H., {Song}, I., \& {Zuckerman}, B. 2007, {\em \apj\/}, {\bf 671},
  616

\bibitem[\protect\astroncite{{Rhee} {\em et~al.\/}}{2008}]{rhee08}
--- 2008, {\em \apj\/}, {\bf 675}, 777

\bibitem[\protect\astroncite{{Rieke} {\em et~al.\/}}{2005}]{rieke05}
{Rieke}, G.~H., {\em et~al.\/} 2005, {\em \apj\/}, {\bf 620}, 1010

\bibitem[\protect\astroncite{{Schlichting} {\em et~al.\/}}{2012}]{sch12}
{Schlichting}, H.~E., {Warren}, P.~H., \& {Yin}, Q.-Z. 2012, {\em \apj\/}, {\bf
  752}, 8

\bibitem[\protect\astroncite{{Schneider} {\em et~al.\/}}{2013}]{schneider13}
{Schneider}, A., {Song}, I., {Melis}, C., {Zuckerman}, B., {Bessell}, M.,
  {Hufford}, T., \& {Hinkley}, S. 2013, {\em \apj\/}, {\bf 777}, 78

\bibitem[\protect\astroncite{{Song} {\em et~al.\/}}{2005}]{song05}
{Song}, I., {Zuckerman}, B., {Weinberger}, A.~J., \& {Becklin}, E.~E. 2005,
  {\em \nat\/}, {\bf 436}, 363

\bibitem[\protect\astroncite{{Stark} \& {Kuchner}}{2009}]{stark09}
{Stark}, C.~C. \& {Kuchner}, M.~J. 2009, {\em \apj\/}, {\bf 707}, 543

\bibitem[\protect\astroncite{{Stewart} \& {Leinhardt}}{2012}]{stewart12}
{Stewart}, S.~T. \& {Leinhardt}, Z.~M. 2012, {\em \apj\/}, {\bf 751}, 32

\bibitem[\protect\astroncite{{Su} {\em et~al.\/}}{2006}]{su06}
{Su}, K.~Y.~L., {\em et~al.\/} 2006, {\em \apj\/}, {\bf 653}, 675

\bibitem[\protect\astroncite{{Tsiganis} {\em et~al.\/}}{2005}]{tsiganis05}
{Tsiganis}, K., {Gomes}, R., {Morbidelli}, A., \& {Levison}, H.~F. 2005, {\em
  \nat\/}, {\bf 435}, 459

\bibitem[\protect\astroncite{{Wyatt}}{2008}]{wyatt08}
{Wyatt}, M.~C. 2008, {\em \araa\/}, {\bf 46}, 339

\bibitem[\protect\astroncite{{Zuckerman} {\em et~al.\/}}{2003}]{zuckerman03}
{Zuckerman}, B., {Koester}, D., {Reid}, I.~N., \& {H{\"u}nsch}, M. 2003, {\em
  \apj\/}, {\bf 596}, 477

\bibitem[\protect\astroncite{{Zuckerman} {\em et~al.\/}}{2010}]{zuckerman10}
{Zuckerman}, B., {Melis}, C., {Klein}, B., {Koester}, D., \& {Jura}, M. 2010,
  {\em \apj\/}, {\bf 722}, 725

\bibitem[\protect\astroncite{{Zuckerman} {\em et~al.\/}}{2012}]{zuckerman12}
{Zuckerman}, B., {Melis}, C., {Rhee}, J.~H., {Schneider}, A., \& {Song}, I.
  2012, {\em \apj\/}, {\bf 752}, 58

\end{thebibliography}
